\documentclass[12pt]{article}
\usepackage{amssymb}
\usepackage{rotating}
\renewcommand{\theequation}{\arabic{section}.\arabic{equation}}

\begin{document}



\def\a{\alpha}
\def\b{\beta}
\def\d{\delta}
\def\e{\epsilon}
\def\g{\gamma}
\def\h{\mathfrak{h}}
\def\k{\kappa}
\def\l{\lambda}
\def\o{\omega}
\def\p{\wp}
\def\r{\rho}
\def\t{\tau}
\def\s{\sigma}
\def\z{\zeta}
\def\x{\xi}
\def\V={{{\bf\rm{V}}}}
 \def\A{{\cal{A}}}
 \def\B{{\cal{B}}}
 \def\C{{\cal{C}}}
 \def\D{{\cal{D}}}
\def\G{\Gamma}
\def\K{{\cal{K}}}
\def\O{\Omega}
\def\R{\bar{R}}
\def\T{{\cal{T}}}
\def\L{\Lambda}
\def\f{E_{\tau,\eta}(sl_2)}
\def\E{E_{\tau,\eta}(sl_n)}
\def\Zb{\mathbb{Z}}
\def\Cb{\mathbb{C}}

\def\R{\overline{R}}

\def\beq{\begin{equation}}
\def\eeq{\end{equation}}
\def\bea{\begin{eqnarray}}
\def\eea{\end{eqnarray}}
\def\ba{\begin{array}}
\def\ea{\end{array}}
\def\no{\nonumber}
\def\le{\langle}
\def\re{\rangle}
\def\lt{\left}
\def\rt{\right}

\newtheorem{Theorem}{Theorem}
\newtheorem{Definition}{Definition}
\newtheorem{Proposition}{Proposition}
\newtheorem{Lemma}{Lemma}
\newtheorem{Corollary}{Corollary}
\newcommand{\proof}[1]{{\bf Proof. }
        #1\begin{flushright}$\Box$\end{flushright}}

\baselineskip=20pt

\newfont{\elevenmib}{cmmib10 scaled\magstep1}
\newcommand{\preprint}{
   \begin{flushleft}
   \end{flushleft}\vspace{-1.3cm}
   \begin{flushright}\normalsize
   \end{flushright}}
\newcommand{\Title}[1]{{\baselineskip=26pt
   \begin{center} \Large \bf #1 \\ \ \\ \end{center}}}
\newcommand{\Author}{\begin{center}
   \large \bf
Guang-Liang Li${}^{a}$, Junpeng Cao${}^{b,c,d}$, Kun Hao${}^{e,f}$,
Fakai Wen${}^{e,f}$, Wen-Li Yang${}^{e,f,g}\footnote{Corresponding
author: wlyang@nwu.edu.cn}$ and Kangjie Shi${}^{e,f}$
 \end{center}}
\newcommand{\Address}{\begin{center}

     ${}^a$Department of Applied Physics, Xian Jiaotong University, Xian 710049, China\\

     ${}^b$Institute of Physics, Chinese Academy of Sciences, Beijing
           100190, China\\
     ${}^c$School of Physical Sciences, University of Chinese Academy of
     Sciences, Beijing, China\\
     ${}^d$Collaborative Innovation Center of Quantum Matter, Beijing,
     China\\
     ${}^e$Institute of Modern Physics, Northwest University,
     Xian 710069, China\\
      ${}^f$Shaanxi Key Laboratory for Theoretical Physics Frontiers, Northwest University,  Xian 710069, China\\
     ${}^g$Beijing Center for Mathematics and Information Interdisciplinary Sciences, Beijing, 100048,  China

   \end{center}}
\newcommand{\Accepted}[1]{\begin{center}
   {\large \sf #1}\\ \vspace{1mm}{\small \sf Accepted for Publication}
   \end{center}}

\preprint
\thispagestyle{empty}
\bigskip\bigskip\bigskip

\Title{Exact solution of the trigonometric $SU(3)$ spin chain with
generic off-diagonal boundary reflections} \Author

\Address
\vspace{1cm}

\begin{abstract}

The nested off-diagonal Bethe ansatz is generalized to study the
quantum spin chain associated with the $SU_q(3)$ $R$-matrix and
generic integrable non-diagonal boundary conditions. By using the
fusion technique, certain closed operator identities among the fused
transfer matrices at the inhomogeneous points are derived. The
corresponding asymptotic behaviors of the transfer matrices and
their values at some special points are given in detail. Based on
the functional analysis, a nested inhomogeneous $T-Q$ relations and
Bethe ansatz equations of the system are obtained. These results can
be naturally generalized to cases related to the $SU_q(n)$ algebra.

\vspace{1truecm} \noindent {\it PACS:} 75.10.Pq, 02.30.Ik, 71.10.Pm

\noindent {\it Keywords}: Spin chain; reflection equation; Bethe
ansatz; $T-Q$ relation
\end{abstract}
\newpage
\section{Introduction}
\label{intro} \setcounter{equation}{0}

Exact solution is a very important issue in studies of statistical
mechanics, condensed matter physics, quantum field theory and
mathematical physics \cite{Bax82,Kor93} since those results can
provide important benchmarks for understanding physical effects in a
variety of systems. The coordinate Bethe ansatz and the algebraic
Bethe ansatz are two powerful methods to obtain the exact solution
of the integrable systems \cite{bax1,Tak79,Skl78,Alc87,Skl88}. With
these methods, many interesting exactly solvable models, such as the
one-dimensional Hubbard model, supersymmetric $t-J$ model,
Heisenberg spin chain and the $\delta$-potential quantum gas model,
were exactly solved. For integrable systems with $U(1)$ symmetry, it
is easy to find a reference state and these conventional Bethe
ansatz can be applied to. Indeed, most of the previous studies focus
on periodic or diagonal open boundary conditions without breaking
the $U(1)$ symmetry. However, there exists another kind of
integrable systems which does not have the $U(1)$ symmetry, such as
the integrable systems with generic off-diagonal boundary
reflections. Because the reference state of this kind of integrable
system is absent, the conventional Bethe ansatz methods are failed.
On the other hand, many interesting phenomena arise in this kind of
systems, such as the topological elementary excitations in the
spin-1/2 torus \cite{Cao1}, spiral phase in the Heisenberg model
with unparallel boundary magnetic field \cite{Cao03} and stochastic
process in non-equilibrium statistical mechanics \cite{Gie05,
Gie05-1,Baj06}. Motivated by these important applications, many
interesting methods such as the q-Onsager algebra
\cite{Bas1,Bas2,Bas3}, the modified algebraic Bethe ansatz
\cite{Bel13,Bel15,Bel15-1,Ava15} and the Sklyanin's separation of
variables (SoV) \cite{Skl92,Fra08,Nic12,Fad14,Kit14} were also
applied to some integrable models without $U(1)$ symmetry. Other
interesting progress can be found in
\cite{Nep04,Yan05,Doi06,Yan06,Nep13}.

Recently, a new approach, i.e., the off-diagonal Bethe ansatz (ODBA)
\cite{Cao1} was proposed to obtain exact solutions of generic
integrable models either with or without $U(1)$ symmetry. Several
long-standing problems were then solved
\cite{Cao13NB875,Cao14NB886,Cao13NB877,Cao14JHEP143,Li14,Zha13,Hao14}
via this method. For comprehensive introduction to this method we
refer the readers to \cite{Wan15}. In order to study the high rank
integrable models, the nested version of ODBA has been  proposed for
the isotropic (or rational) models \cite{Cao14JHEP143}. In this
paper, we study the anisotropic rank-2 spin model with generic
integrable boundary conditions. Here the $R$-matrix is the
trigonometric one associated with  the $SU_q(3)$ algebra and the
boundary reflection matrices are the most generic reflection
matrices which have non-vanishing off-diagonal elements. Because the
off-diagonal elements of the reflection matrices break the $U(1)$
symmetry, the exact solution of the system has been missing even its
integrability was known for many years ago. By using the fusion
technique and nested ODBA, we successfully obtain the closed
operator identities, the values at the special points and the
asymptotic behaviors. Based on them, we construct the nested
inhomogeneous $T-Q$ relation and obtain the eigenvalue of the
transfer matrix thus the energy spectrum of the system.  These
results can be generalized to multiple components spin chains
related to more higher rank algebra cases.

The paper is organized as follows. Section 2 is the general
description of the model. The $SU_q(3)$ $R$-matrix and corresponding
generic integral non-diagonal boundary reflection matrices are
introduced. In Section 3, by using the fusion technique, we derive
the closed operator identities for the fused transfer matrices and
the quantum determinant. The asymptotic behaviors of the fused
transfer matrix and their values at special points are also
obtained. In section 4, we list some necessary functional relations
which are used to determine the eigenvalues. Section 5 is devoted to
the construction of the nested inhomogeneous $T-Q$ relation and the
Bethe ansatz equations. In section 6, we summarize our results and
give some discussions. Some results related to the other types of
the general off-diagonal  boundary reflections are given in
Appendix.


\section{The model}
\label{XXZ} \setcounter{equation}{0}

Throughout, ${\rm\bf V}$ denotes a three-dimensional linear space
and let $\{|i\rangle, i=1,2,3\}$ be an orthonormal basis of it. We
shall adopt the standard notations. For any matrix $A\in {\rm
End}({\rm\bf V})$, $A_j$ is an embedding operator in the tensor
space ${\rm\bf V}\otimes {\rm\bf V}\otimes\cdots$, which acts as $A$
on the $j$-th space and as identity on the other factor spaces. For
$B\in {\rm End}({\rm\bf V}\otimes {\rm\bf V})$, $B_{ij}$ is an
embedding operator of $B$ in the tensor space, which acts as
identity on the factor spaces except for the $i$-th and $j$-th ones.

The $R$-matrix $R(u)\in {\rm End}({\rm\bf V}\otimes {\rm\bf V})$
used in this paper is the trigonometric one associated with the
$SU_q(3)$ algebra, which was first proposed by Perk and Shultz
\cite{Perk81} and further studied in
\cite{Perk83,Schu83,Perk06,Baz85,Jim86}, \bea
 R_{12}(u)=\left(\begin{array}{r|r|r}{\begin{array}{rrr}a(u)&&\\&b(u)&\\&&b(u)\end{array}}
           &{\begin{array}{lll}&&\\c(u)&{\,\,\,\,\,\,\,\,\,\,}&{\,\,\,\,\,\,\,\,\,\,}\\{\,\,\,\,\,\,}&{\,\,\,\,\,\,}&{\,\,\,\,\,\,}\end{array}}
           &{\begin{array}{lll}&&\\&&\\c(u)&{\,\,\,\,\,\,\,\,\,\,}&{\,\,\,\,\,\,\,\,\,\,}\end{array}}\\[12pt]
 \hline {\begin{array}{rrr}&d(u)&\\&&{\,\,\,\,\,\,\,\,\,\,\,}\\&&\end{array}}&
           {\begin{array}{ccc}b(u)&&\\&a(u)&\\&&b(u)\end{array}}
           &{\begin{array}{lll}&&\\{\,\,\,\,\,\,\,\,\,\,}&{\,\,\,\,\,\,\,\,\,\,}&{\,\,\,\,\,\,\,\,\,\,}\\&c(u)&{\,\,\,\,\,\,\,\,\,\,}\end{array}}\\[12pt]
 \hline {\begin{array}{ccc}&&d(u)\\&&\\&&\end{array}}
           &{\begin{array}{ccc}&&\\&&d(u)\\&&\end{array}}
           &{\begin{array}{ccc}b(u)&&\\&b(u)&\\&&a(u)\end{array}} \end{array}\right),\label{R-matrix}
\eea \noindent where the matrix elements are \bea
&&a(u)=\sinh(u+\eta), \quad
b(u)=\sinh(u), \\[4pt]
&&c(u)=e^{u}\sinh\eta, \quad d(u)=e^{-{u}}\sinh \eta.
\label{R-element-2} \eea
 The
$R$-matrix satisfies the quantum Yang-Baxter equation (QYBE) \bea
R_{12}(u_1-u_2)R_{13}(u_1-u_3)R_{23}(u_2-u_3)
=R_{23}(u_2-u_3)R_{13}(u_1-u_3)R_{12}(u_1-u_2),\label{QYB}\eea and
possesses the following   properties, \bea &&\hspace{-1.5cm}\mbox{
Initial
condition}:\,R_{12}(0)= \sinh\eta P_{12},\label{Int-R}\\
&&\hspace{-1.5cm}\mbox{ Unitarity
relation}:\,R_{12}(u)R_{21}(-u)= \rho_1(u)\,\times {\rm id},\label{Unitarity}\\
&&\hspace{-1.5cm}\mbox{ Crossing Unitarity
relation}:\,R_{12}^{t_1}(u)\,{\cal M}_1\,R_{21}^{t_1}(-u-3\eta)\,{\cal M}_1^{-1}=\rho_2(u)\,\times {\rm id},\label{crossing-unitarity} \\[6pt]
&&\hspace{-1.5cm}\mbox{ PT-symmetry}:\,R_{21}(u)=R^{t_1\,t_2}_{12}(u),\label{PT}\\
&&\hspace{-1.5cm}\mbox{ Periodicity}: \,
R_{12}(u+i\pi)=-R_{12}(u).\label{Periodic} \eea Here
$R_{21}(u)=P_{12}R_{12}(u)P_{12}$ with $P_{12}$ being the usual
permutation operator and $t_i$ denotes transposition in the $i$-th
space. The functions $\rho_1(u)$, $\rho_2(u)$ and the crossing
matrix ${\cal M}$ are given by \bea
\rho_1(u)&=&-\sinh({u}-\eta)\sinh({u}+\eta),\label{p-1-function}\\[4pt]
\rho_2(u)&=&-\sinh({u})\sinh({u}+3\eta),\label{p-2-function}\\[4pt]
{\cal
M}&=&\left(\begin{array}{ccc}e^{4\eta}&&\\&e^{2\eta}&\\&&1\end{array}\right).\label{V-matrix}
\eea It is easy to check that the $R$-matrix (\ref{R-matrix}) also
has the following properties \bea {\cal M}_1\,{\cal
M}_2\,R_{12}(u)\,{\cal M}_1^{-1}\,{\cal M}_2^{-1}=R_{12}(u).\label{Invariant} \eea

Let us introduce the reflection matrix $K^-(u)$ and its dual one
$K^+(u)$. The former satisfies the reflection equation (RE)
 \bea &&R_{12}(u_1-u_2)K^-_1(u_1)R_{21}(u_1+u_2)K^-_2(u_2)\no\\
 &&~~~~~~=
K^-_2(u_2)R_{12}(u_1+u_2)K^-_1(u_1)R_{21}(u_1-u_2),\label{RE-V}\eea
and the latter satisfies the dual RE \bea
&&R_{12}(u_2-u_1)K^+_1(u_1){\cal M}_1^{-1}R_{21}(-u_1-u_2-3\eta){\cal M}_1K^+_2(u_2)\no\\[6pt]
&&~~~~~~= K^+_2(u_2){\cal M}_2^{-1}R_{12}(-u_1-u_2-3\eta){\cal
M}_2K^+_1(u_1)R_{21}(u_2-u_1). \label{DRE-V}\eea In this paper we
consider the generic non-diagonal $K$-matrices $K^-(u)$ found in
\cite{Shik,Shik2,Shik3}. There are three kinds of reflecting
$K$-matrix: \bea
({\rm{I}}): K^-(u)=\left(\begin{array}{ccc}e^{u}\sinh(\zeta-u)+ce^{2u}\sinh (2u)&0&0\\[6pt]
0&e^{u}\sinh(\zeta-u)&c_1\sinh (2u)\\[6pt]
0&c_2\sinh (2u)&e^{-u}\sinh(\zeta+u)\end{array}\right),
\label{K-matrix-1} \eea with the constraint \bea
c^2=c_1c_2+ce^{\zeta}. \no\eea Thus the four boundary parameters
$c$, $c_1$ $c_2$ and $\zeta$ are not independent with each other.
\bea
({\rm{II}}): K^-(u)=\left(\begin{array}{ccc}e^{u}\sinh(\zeta-u)&0&c_1\sinh (2u)\\[6pt]
0&e^{u}\sinh(\zeta-u)+c\sinh (2u)&0\\[6pt]
c_2\sinh (2u)&0&e^{-u}\sinh(\zeta+u)\end{array}\right),
\label{K-matrix-2} \eea with the constraint \bea
c^2=c_1c_2+ce^{-\zeta}. \no \eea \bea { ({\rm{III}}):
K^-(u)=\left(\begin{array}{ccc}e^{u}\sinh(\zeta-u)&c_1\sinh (2u)&0\\[6pt]
c_2\sinh (2u)&e^{-u}\sinh(\zeta+u)&0\\[6pt]
0&0&e^{-u}\sinh(\zeta+u)+ce^{-2u}\sinh (2u)\end{array}\right),}
\label{K-matrix-3} \eea with the constraint \bea
c^2=c_1c_2-ce^{\zeta}. \no\eea The dual non-diagonal reflection
matrix $K^+(u)$ is given by \bea K^+(u)={\cal
M}K^-(-u-3\eta/2)\left|_{(\zeta,c,c_1,c_2)\rightarrow
(\zeta',c',c_1',c_2')}\right..\label{K-matrix-4} \eea

In order to construct the model's Hamiltonian of the system, we
first introduce the ``row-to-row" (or one-row) monodromy matrices
$T_0(u)$ and $\hat{T}_0(u)$ \bea
T_0(u)&=&R_{0N}(u-\theta_N)R_{0\,N-1}(u-\theta_{N-1})\cdots
R_{01}(u-\theta_1),\label{Mon-V-1}\\
\hat{T}_0(u)&=&R_{10}(u+\theta_1)R_{20}(u+\theta_{2})\cdots
R_{N0}(u+\theta_N),\label{Mon-V-2} \eea where $\{\theta_j, j=1,
\cdots, N\}$ are the inhomogeneous parameters and $N$ is the number
of sites. The one-row monodromy matrices are the $3\times 3$
matrices in the auxillary space $0$ and their elements act on the
quantum space ${\rm\bf V}^{\otimes N}$. For the system with open
boundaries, we need to define the double-row monodromy matrix
$\mathbb{T}_0(u)$ \bea
  \mathbb{T}_0(u)=T_0(u)K^-_0(u)\hat{T}_0(u).
  \label{Mon-V-0}
\eea Then the transfer matrix of the system is constructed as
\cite{Skl88} \bea
t(u)=tr_0\{K^+_0(u)\mathbb{T}_0(u)\}.\label{trans}\eea From the QYBE
(\ref{QYB}), RE (\ref{RE-V}) and dual RE (\ref{DRE-V}), one can
prove that the transfer matrices with different spectral parameters
commute with each other, $[t(u), t(v)]=0$. Therefore, $t(u)$ serves
as the generating functional of all the conserved quantities of the
system. The model Hamiltonian can be constructed by taking the
derivative of the logarithm of the transfer matrix of the system
\begin{eqnarray}
&&H=\sinh\eta \frac{\partial \ln t(u)}{\partial
u}|_{u=0,\{\theta_j\}=0}. \label{hamimi}
\end{eqnarray}


\section{Fusion}

\label{T-QR} \setcounter{equation}{0}

Following \cite{Cao14JHEP143}, we apply the fusion technique
\cite{Kar79,Mez92,Zho96} to study the present model. The fusion
procedure will lead to the desired operator identities to determine
the spectrum of the transfer matrix $t(u)$ given by (\ref{trans}).
For this purpose, let us introduce the following vectors in the
tensor space ${\rm\bf V}\otimes {\rm\bf V}$ similarly as \cite{Hao14}
\bea
|\Phi^{(1)}_{12}\rangle&=&\frac{1}{\sqrt{2e^{\eta}\cosh \eta}}
(|1,2\rangle- e^{\eta}|2,1\rangle), \no\\[4pt]
|\Phi^{(2)}_{12}\rangle&=&\frac{1}{\sqrt{2e^{\eta}\cosh \eta}}
(|1,3\rangle- e^{\eta}|3,1\rangle), \\[4pt]
|\Phi^{(3)}_{12}\rangle&=&\frac{1}{\sqrt{2e^{\eta}\cosh \eta}}
(|2,3\rangle- e^{\eta}|3,2\rangle), \no \label{Vector-01} \eea in
the tensor space ${\rm\bf V}\otimes {\rm\bf V}$ and \bea
|\Phi_{123}\rangle&=&\frac{1}{\sqrt{2e^{3\eta}(2\cosh \eta+\cosh
3\eta)}}
(|1,2,3\rangle- e^{\eta}|1,3,2\rangle+e^{2\eta}|3,1,2\rangle\no\\[6pt]
&&-e^{\eta}|2,1,3\rangle+e^{2\eta}|2,3,1\rangle-e^{3\eta}|3,2,1\rangle),\label{Vector-02}
\eea in the tensor space ${\rm\bf V}\otimes {\rm\bf V}\otimes
{\rm\bf V}$. The associated projectors\footnote{We note
that in contrast to most of rational models, here $P^{-}_{12}\neq
P^{-}_{21}$. Therefore, the orders of sub-indices in (\ref{P-matrix-1-1})-(\ref{P-matrix-4}) are crucial (c.f. \cite{Cao14JHEP143}).    } are \bea
&&P_{12}^{-}=|\Phi^{(1)}_{12}\rangle\langle\Phi^{(1)}_{12}|+|\Phi^{(2)}_{12}\rangle\langle\Phi^{(2)}_{12}|
+|\Phi^{(3)}_{12}\rangle\langle\Phi^{(3)}_{12}|,\\[4pt]
&&P_{123}^{-}=|\Phi_{123}\rangle\langle\Phi_{123}|. \eea Direct
calculation shows that the $R$-matrix given by (\ref{R-matrix}) at
some degenerate points are proportional to the projectors, \bea
R_{12}(-\eta)=P^{-}_{12}\times S_{12},\quad
R_{12}(-\eta)R_{13}(-2\eta)R_{23}(-\eta)=P_{123}^{-}\times
S_{123},\label{Fusion} \eea where the diagonal matrices $S_{12}$ and
$S_{123}$ are given by \bea &&S_{12}=-\sinh 2\eta\times Diag[
1,e^{\eta},e^{\eta},e^{-\eta},1,e^{\eta},e^{-\eta},e^{-\eta},1],\\[4pt]
&&S_{123}=-2\sinh 2\eta \sinh^2 \eta(2\cosh\eta+\cosh 3\eta)\times
Diag[1,1,1,1,1,e^{3\eta},1,e^{\eta},1,1,1,e^{\eta},\no\\
&&
\hspace{1cm}1,1,1,e^{-\eta},1,1,1,e^{-\eta},1,e^{-3\eta},1,1,1,1,1].
\eea

The fused transfer matrices are defined as \bea t_m(u)=tr_{12\cdots
m}\{ K^+_{<12\cdots m>}(u)T_{<12\cdots m>}(u)K^-_{<12\cdots
m>}(u)\hat{T}_{<12\cdots m>}(u)\}, \quad m=1, 2, 3,
\label{P-matrix-1} \eea where \bea &&K^{+}_{12\cdots
m}(u)=K^+_{<2\cdots
m>}(u-\eta){\cal M}_2^{-1}R_{1m}(-2u+(m-1)\eta-3\eta)\no\\[4pt]
&& \qquad \qquad \times R_{1m-1}(-2u+(m-2)\eta-3\eta)\cdots
R_{12}(-2u+\eta-3\eta){\cal
M}_2K^{+}_1(u), \label{P-matrix-1-1}\\[4pt]
&&K^+_{<12\cdots m>}(u)=P^-_{12\cdots m}K^{+}_{12\cdots
m}(u)P^-_{mm-1\cdots 1}, \\[4pt]
&&K^{-}_{12\cdots
m}(u)=K^-_1(u)R_{21}(2u-\eta)\no\\[4pt]
&&\qquad \qquad \qquad \times R_{31}(2u-2\eta)\cdots
R_{m1}(2u-(m-1)\eta)K^{-}_{<2\cdots
m>}(u-\eta), \label{P-matrix-2} \\[4pt]
&&K^-_{<12\cdots m>}(u)=P^-_{mm-1\cdots 1}K^{-}_{12\cdots
m}(u)P^-_{12\cdots m}, \label{P-matrix-5} \\[4pt]
&&T_{<12\cdots m>}(u)=P^-_{mm-1\cdots 1}T_1(u)T_2(u-\eta)\cdots
T_m(u-(m-1)\eta)P^-_{mm-1\cdots 1}, \label{P-matrix-3} \\[4pt]
&&\hat{T}_{<12\cdots m>}(u)=P^-_{12\cdots
m}\hat{T}_1(u)\hat{T}_2(u-\eta)\cdots
\hat{T}_m(u-(m-1)\eta)P^-_{12\cdots m}, \label{P-matrix-4}
 \eea
and the notation $t_1(u)=t(u)$ is used. By repeatedly using the QYBE
(\ref{QYB}), the RE (\ref{RE-V}), the dual RE (\ref{DRE-V}) and the
definition (\ref{P-matrix-1}), one can prove that all these fused
transfer matrices are commutative with each other \bea [t_m(u),
t_k(v)]=0, \quad m, k=1, 2, 3. \label{commu-trans-su3} \eea Thus
they have the common eigenstates. Furthermore, we find that the
transfer matrix given by (\ref{P-matrix-1}) satisfies the following
operator production identities \bea
&&t(\pm\theta_j)t_m(\pm\theta_j-\eta)={t_{m+1}(\pm\theta_j)}
            {\prod_{k=1}^m\rho^{-1}_2(\pm2\theta_j-k\eta)},\,\,  j=1, \ldots, N,\,\, m=1, 2, \label{Id-1}\\[4pt]
&&t_2(\pm\theta_j+\eta)=0,\quad j=1,\cdots,N.\label{Extra-1}
\eea We note that the fused transfer matrix $t_3(u)$ equals to its
 quantum determinant multiplying the unity matrix. Thus the
operators production identities (\ref{Id-1}) are closed. The
explicit form of the fused transfer matrix $t_3(u)$ reads
\begin{eqnarray} && t_3(u)
=\Delta_q(u)\times {\rm id}=\Delta_q\{T(u)\}\Delta_q\{\hat
T(u)\}\Delta_q\{K^+(u)\}\Delta_q\{K^-(u)\} \times {\rm id},
\label{1109-1}
\end{eqnarray}
where $\Delta_q\{T(u)\}$, $\Delta_q\{\hat T(u)\}$,
$\Delta_q\{K^+(u)\}$ and $\Delta_q\{K^-(u)\}$ are the quantum
determinants of the matrices $T(u)$, $\hat{T}(u)$, $K^+(u)$ and
$K^-(u)$, respectively. The quantum determinants of the one-row
monodromy matrices are \bea \Delta_q\{T(u)\} =
\prod_{l=1}^N\sinh(u-\theta_l+\eta)
\sinh(u-\theta_l-\eta)\sinh(u-\theta_l-2\eta), \\
\Delta_q\{\hat T(u)\}= \prod_{l=1}^N\sinh(u+\theta_l+\eta)
\sinh(u+\theta_l-\eta)\sinh(u+\theta_l-2\eta). \eea The quantum
determinant of the reflecting matrix (I) given by (\ref{K-matrix-1}) is
\bea \Delta_q\{K^-(u)\}&=&
 -(e^{u-\eta}\sinh(\zeta-u+\eta)+ce^{2u-2\eta}\sinh
(2u-2\eta))\no\\
&&\times(\sinh(\zeta-u)\sinh(\zeta+u)-c_1c_2\sinh
(2u)\sinh(2u))\no\\
&&\times\sinh(2u-2\eta)\sinh(2u-3\eta)\sinh(2u-4\eta)\no\\
 &=&
-(e^{u-\eta}\sinh(\zeta-u+\eta)+ce^{2u-2\eta}\sinh
(2u-2\eta))\no\\
&&\times(e^{u}\sinh(\zeta-u)+ce^{2u}\sinh
(2u))\no\\
&&\times(e^{-u}\sinh(\zeta+u)-ce^{-2u}\sinh
(2u))\no\\
&&\times\sinh(2u-2\eta)\sinh(2u-3\eta)\sinh(2u-4\eta).\label{p-3-function-1}
\eea The quantum determinant of the reflecting matrix (II) given by (\ref{K-matrix-2}) is
\bea \Delta_q\{K^-(u)\} &=&
-(e^{u-\eta}\sinh(\zeta-u+\eta)+c\sinh
(2u-2\eta))\no\\
&&\times(\sinh(\zeta-u)\sinh(\zeta+u)-c_1c_2\sinh
(2u)\sinh(2u))\no\\
&&\times\sinh(2u-2\eta)\sinh(2u-3\eta)\sinh(2u-4\eta)\no\\
 &=&
-(e^{u-\eta}\sinh(\zeta-u+\eta)+c\sinh
(2u-2\eta))\no\\
&&\times(e^{u}\sinh(\zeta-u)+c\sinh
(2u))\no\\
&&\times(e^{-u}\sinh(\zeta+u)-c\sinh
(2u))\no\\
&&\times\sinh(2u-2\eta)\sinh(2u-3\eta)\sinh(2u-4\eta).
\label{p-3-function-2} \eea The quantum determinant of the
reflecting matrix (III) given by (\ref{K-matrix-3}) is \bea
\Delta_q\{K^-(u)\}&=&
 -(e^{-u+\eta}\sinh(\zeta+u-\eta)+ce^{-2u+2\eta}\sinh
(2u-2\eta))\no\\
&&\times(\sinh(\zeta-u)\sinh(\zeta+u)-c_1c_2\sinh
(2u)\sinh(2u))\no\\
&&\times\sinh(2u-2\eta)\sinh(2u-3\eta)\sinh(2u-4\eta)\no\\
 &=&
-(e^{-u+\eta}\sinh(\zeta+u-\eta)+ce^{-2u+2\eta}\sinh
(2u-2\eta))\no\\
&&\times(e^{u}\sinh(\zeta-u)-ce^{2u}\sinh
(2u))\no\\
&&\times(e^{-u}\sinh(\zeta+u)+ce^{-2u}\sinh
(2u))\no\\
&&\times\sinh(2u-2\eta)\sinh(2u-3\eta)\sinh(2u-4\eta).\label{p-3-function-3}
\eea The quantum determinant of the dual reflecting matrices
$K^+(u)$ can be obtained by the mapping \bea
\Delta_q\{K^+(u)\}=e^{6\eta}\Delta_q\{K^-(-u+\frac{\eta}{2})\}|_{(c,
\zeta\rightarrow c', \zeta')}. \no\eea
Then the equation (\ref{1109-1}) can be proved
easily based on the facts
\bea &&T_{<123>}(u)=\Delta_q\{T(u)\}P^-_{321},\label{Fused-R-1}\\[6pt]
&&\hat{T}_{<123>}(u)=\Delta_q\{\hat
T(u)\} P^-_{123},\label{Fused-R-2}\\[6pt]
&&K^-_{<123>}(u)=\Delta_q\{K^-(u)\}|\Phi_{321}\rangle\langle\Phi_{123}|,\label{Fused-R-3}\\[6pt]
&&K^+_{<123>}(u)=\Delta_q\{K^+(u)\}|\Phi_{123}\rangle\langle\Phi_{321}|.\label{Fused-R-4}
\eea

Form the definition of fused transfer matrices (\ref{P-matrix-1}),
the corresponding asymptotic behaviors can be calculated directly.
Obviously, different reflection parameters will give different
asymptotic behaviors. Without losing the generality, we consider the
case corresponding to the reflection matrices $K^{\pm}(u)$ given by
(\ref{K-matrix-1}) and (\ref{K-matrix-4}) and the details for the
results for the other cases will be presented in Appendix.  Then the
asymptotic behaviors read \bea
t_1(u)|_{u\rightarrow +\infty}&=&-\frac{1}{4^{N+1}}e^{(2N+4)u+3\eta}\left[c\frac{c_1'c_2'}{c'}e^{\eta}e^{2\eta {\cal{Q}}^{(1)}}\rt.\no\\
&&+\lt.(c_1c_2'+c_1'c_2e^{2\eta})e^{N\eta}e^{-\eta {\cal{Q}}^{(1)}}\right]+\cdots, \label{Id-2}\\[4pt]
t_1(u)|_{u\rightarrow -\infty}&=&-\frac{1}{4^{N+1}}e^{-(2N+4)u-3\eta}\left[c'\frac{c_1c_2}{c}e^{\eta}e^{-2\eta {\cal{Q}}^{(1)}}\rt.\no\\
&&+\lt.(c_1c_2'\hspace{-0.08truecm}+\hspace{-0.08truecm}c_1'c_2e^{2\eta})
e^{-N\eta} e^{\eta {\cal{Q}}^{(1)}}\right]\hspace{-0.08truecm}+\hspace{-0.08truecm}\cdots, \label{Id-3}\\[4pt]
t_2(u)|_{u\rightarrow +\infty}&=&-\frac{1}{4^{2N+3}}e^{4(N+3)u+4\eta}c\frac{c_1'c_2'}{c'}\left[c'\frac{c_1c_2}{c}e^{\eta}
e^{-2\eta {\cal{Q}}^{(1)}}\right.\no\\[4pt]
&&+\left.(c_1c_2'+c_1'c_2e^{2\eta})e^{-N\eta}e^{\eta {\cal{Q}}^{(1)}}\right]+\cdots, \label{Id-4}\\[4pt]
t_2(u)|_{u\rightarrow -\infty}&=&-\frac{1}{4^{2N+3}}e^{-4(N+3)u-2\eta}c'\frac{c_1c_2}{c}\left[c\frac{c_1'c_2'}{c'}e^{\eta}e^{2\eta {\cal{Q}}^{(1)}}\right.\no\\[4pt]
&&+\left.(c_1c_2'+c_1'c_2e^{2\eta})e^{N\eta}e^{-\eta {\cal{Q}}^{(1)}}\right]+\cdots,
\label{Id-5} \eea where the operator ${\cal{Q}}^{(1)}$ is
\bea
{\cal{Q}}^{(1)}=\sum_{l=1}^N E^{11}_l,\quad E^{11}=\left(\begin{array}{ccc}1&0&0\\0&0&0\\0&0&0\end{array}\rt). \label{Q-operator}
\eea
In the derivation, the relations $c(c-e^{\zeta})=c_1c_2$ and
$c'(c'-e^{\zeta'})=c_1'c_2'$ are used. It is remarked that the non-diagonal K-matrices (given by (\ref{K-matrix-1}) and (\ref{K-matrix-4})) only break two of the original three $U(1)$-symmetries for
the diagonal K-matrices or periodical case, and that the system still has a remaining  $U(1)$ -symmetry which
is generated by the operator ${\cal{Q}}^{(1)}$.

The fused transfer matrices $t_m(u)$ have other useful properties.
For example, their values at some special points can be calculated
directly by using the properties of the $R$-matrix and the
reflection matrices $K^{\pm}$. We list them in the following
\bea
&&t_m(u+i\pi)=t_m(u),\quad m=1,2,\label{t1-1-0}\\
&&t(0)= \prod_{l=1}^N\rho_1
(-\theta_l)tr \{K^+(0)\}K^-(0)\,\times{\rm id},\label{t1-1} \\[4pt]
&&t(\frac{i \pi}{2})=  (-1)^N\prod_{l=1}^N\rho_1
(-\theta_l+\frac{i \pi}{2})tr \{K^+(\frac{i \pi}{2})\}K^-(\frac{i \pi}{2})\times{\rm id},\label{t1-11} \\[4pt]
&&t(-\frac{3\eta}{2})=\prod_{l=1}^N
\rho_2(-\theta_l-\frac{3\eta}{2})tr
\{K^-(-\frac{3\eta}{2}){\cal M}\}{\cal M}^{-1}K^+(-\frac{3\eta}{2})\times {\rm id}, \label{t1-2}\\[4pt]
&&t(-\frac{3\eta}{2}+\frac{i \pi}{2})=(-1)^N\prod_{l=1}^N
\rho_2(-\theta_l-\frac{3\eta}{2}-\frac{i \pi}{2})tr
\{K^-(-\frac{3\eta}{2}+\frac{i \pi}{2}){\cal M}\}\no\\[4pt]
&&\qquad\qquad \qquad \quad \times {\cal M}^{-1}K^+(-\frac{3\eta}{2}+\frac{i \pi}{2})\times {\rm id}, \label{t1-22}\\[4pt]
&& t_2(\frac{\eta}{2})=tr_{12}\left\{ P_{12}^-
K_{2}^+(-\frac{\eta}{2}){\cal M}_2^{-1}R_{12}(-3\eta){\cal M}_2K_{1}^+(\frac{\eta}{2})R_{12}(0)P_{12}^-\right\}\rho_K^-(\frac{\eta}{2})
\no \\[4pt]
&&\qquad\qquad \qquad \quad \times \prod_{l=1}^N
\rho_1(\frac\eta2-\theta_l)\rho_1(-\frac\eta2-\theta_l)
\times {\rm id}, \label{t2-2} \\[4pt]
&& t_2(\frac{\eta}{2}+\frac{i \pi}{2})=tr_{12}\left\{ P_{12}^-
K_{2}^+(-\frac{\eta}{2}+\frac{i \pi
}{2}){\cal M}_2^{-1}R_{12}(-3\eta){\cal M}_2K_{1}^+(\frac{\eta}{2}+\frac{i \pi
}{2})R_{12}(0)P_{12}^-\right\}
\no \\[4pt]
&&\qquad \qquad \qquad \quad \times \rho_K^-(\frac{\eta}{2}+\frac{i
\pi }{2})\prod_{l=1}^N \rho_1(\frac\eta2-\theta_l+\frac{i \pi
}{2})\rho_1(-\frac\eta2-\theta_l+\frac{i \pi }{2})
\times {\rm id}, \label{t2-22} \\[4pt]
&& t_2(-\eta)=tr_{12}\left\{
P_{12}^-R_{12}(0)K_{1}^-(-\eta)R_{21}(-3\eta)K_{2}^-(-2\eta){\cal M}_1{\cal M}_2P_{12}^-\right\}\rho_K^+(\frac{\eta}{2})
\no \\[4pt]
&&\qquad\qquad \qquad \quad \times \prod_{l=1}^N
\rho_2(-\theta_l-2\eta)\rho_2(-\theta_l-\eta) \,\times {\rm id},
\label{t2-3}\\[4pt]
&& t_2(-\eta+\frac{i \pi}{2})=tr_{12}\left\{
P_{12}^-R_{12}(0)K_{1}^-(-\eta+\frac{i \pi
}{2})R_{21}(-3\eta)K_{2}^-(-2\eta+\frac{i \pi
}{2}){\cal M}_1{\cal M}_2P_{12}^-\right\}
\no \\[4pt]
&&\qquad\qquad \qquad \quad \times \rho_K^+(\frac{\eta}{2}+\frac{i
\pi }{2})\prod_{l=1}^N \rho_2(-\theta_l-2\eta+\frac{i \pi
}{2})\rho_2(-\theta_l-\eta+\frac{i \pi }{2})\times {\rm id},
\label{t2-33}\\
&& t_2(0)= b(-\eta)b(-2\eta)K^-(0)\prod_{l=1}^N
\rho_1(-\theta_l)tr \{K^+(0)\}\,  t(-\eta),\label{t2-11} \\[4pt]
&& t_2(\frac{i \pi }{2})= b(-\eta)b(-2\eta)K^-(\frac{i \pi
}{2})(-1)^N\prod_{l=1}^N
\rho_1(\frac{i \pi }{2}-\theta_l)tr \{K^+(\frac{i \pi }{2})\}\,  t(-\eta+\frac{i\pi}{2}),\label{t2-1} \\[4pt]
&& t_2(-\frac{\eta}{2})=b(-\eta)b(-2\eta){\cal M}^{-1}K^+(-\frac{3\eta}{2})
\prod_{l=1}^N \rho_2(-\theta_l-\frac{3\eta}{2})tr
\{K^-(-\frac32\eta){\cal M}\} \,t(-\frac{\eta}{2}),\label{t2-4} \\[4pt]
&& t_2(-\frac{\eta}{2}+\frac{i \pi
}{2})=b(-\eta)b(-2\eta){\cal M}^{-1}K^+(-\frac{3\eta}{2}+\frac{i \pi }{2})
(-1)^N\prod_{l=1}^N \rho_2(-\theta_l-\frac{3\eta}{2}+\frac{i \pi
}{2})\no\\[4pt]
&&\qquad\qquad \qquad \quad \times tr \{K^-(-\frac{3\eta}{2}+\frac{i
\pi }{2}){\cal M}\}t(-\frac{\eta}{2}+\frac{i \pi }{2}),\label{t2-44}
\\ [4pt]&& t_2(\eta)=t_2(\eta+\frac{i \pi
}{2})=t_2(-\frac{3\eta}{2})=t_2(-\frac{3\eta}{2}+\frac{i \pi
}{2})=0, \label{ttt2}
\end{eqnarray}
where the notations $\rho_K^-(u)$ and $\rho_K^+(u)$ are defined as
\bea &&\rho_K^-(u)=K^-(u)K^-(-u), \quad
\rho_K^+(u)=\rho_K^-(u)|_{c,\zeta,c_1,c_2\rightarrow
c',\zeta',c_1',c_2'} . \eea In the derivation, we have used the
relations
\bea &&K^{\pm}(u+\pi i)=K^{\pm}(u), \quad R(u+\pi i)=-R(u), \\[4pt]
&&T_0(u)\hat{T}_0(-u)=\prod_{l=1}^N\rho_1(u-\theta_l)\times {\rm id}, \\
&&T^{t_0}_0(u){\cal M}_0\hat{T}^{t_0}_0(-u-3\eta){\cal M}_0^{-1}=\prod_{l=1}^N\rho_2(u-\theta_l)\times
{\rm id}. \label{T-Relation}
 \eea


\section{Functional relations}
\label{func} \setcounter{equation}{0}

Because the fused transfer matrices $t_m(u)$ commute with each
other, they have the common eigenstates. Let $|\Psi\rangle$ be a
common eigenstate of $t_m(u)$, which dose not depend upon $u$, with
the eigenvalues $\Lambda_m(u)$, \bea
t_m(u)|\Psi\rangle=\Lambda_m(u)|\Psi\rangle,\quad m=1,2,3.\no \eea
Again, we use the notation $\Lambda_1(u)=\Lambda(u)$, which
represents the eigenvalue of transfer matrix $t(u)$ given by
(\ref{trans}). The $\Lambda(u)$, as an entire function of $u$, is a
trigonometric polynomial of degree $2N +4$, which can be completely
determined by $2N + 5$ conditions. The $\Lambda_2(u)$, as an entire
function of $u$, is a trigonometric polynomial of degree $4N +12$,
which can be completely determined by $2N + 13$ conditions\footnote{It is noted that the relations (\ref{Extra-1}) give the other $2N$ conditions. }.

From the operator production identities (\ref{Id-1}) and (\ref{t1-1-0}), we have
\bea
&&\hspace{-0.8truecm}\Lambda_m(u+i\pi)=\Lambda_m(u),\quad m=1,2,\label{Ld-1-0}\\
&&\hspace{-0.8truecm}\Lambda(\pm\theta_j)\Lambda_m(\pm\theta_j-\eta)={\Lambda_{m+1}(\pm\theta_j)}
            {\prod_{k=1}^m\rho^{-1}_2(\pm2\theta_j-k\eta)}, ~~  j=1,\ldots,
N, ~~ m=1,2, \label{Ld-1}\\[4pt]
&&\hspace{-0.8truecm}\Lambda_2(\pm\theta_j+\eta)=0, \quad  j=1,\ldots, N,\label{Ld-1-2}\\[4pt]
&&\hspace{-0.8truecm}\Lambda_3(u)=\Delta_q(u).\label{Ld-1-3} \eea

The values of $\Lambda(u)$ at the special points \bea 0, \quad \frac
{ i \pi}{2}, \quad  -\frac{3\eta}{2}, \quad -\frac{3\eta}{2}+\frac{i
\pi}{2} ,  \label{LLd-1} \eea should be the same as those given by
(\ref{t1-1})-(\ref{t1-22}) of the transfer matrix $t(u)$. At the same time,
the values of $\Lambda_2(u)$ at the special points
\bea && 0, \quad
\frac { i \pi}{2}, \quad \frac{\eta}{2}, \quad \frac{\eta}{2}+\frac
{ i
\pi}{2}, \quad  \eta, \quad \eta+\frac { i \pi}{2},  \nonumber  \\
&& -\frac{\eta}{2}, \quad -\frac{\eta}{2}+\frac { i \pi}{2}, \quad -
\eta, \quad - \eta+\frac { i \pi}{2}, \quad -\frac{3\eta}{2}, \quad
-\frac{3\eta}{2}+\frac { i \pi}{2},  \label{LLd-2} \eea should be
the same as those given by (\ref{t2-2})-(\ref{ttt2}) of the fused
transfer matrix $t_2(u)$.

The asymptotic behaviors of $\Lambda_m(u)$ can be obtained by acting
the operators in (\ref{Id-2})-(\ref{Id-5}) on the corresponding
eigenstates. The asymptotic behaviors (\ref{Id-2})-(\ref{Id-5})
allows us to decompose the whole Hilbert space $\cal{H}$ into $N+1$
subspaces, i.e., ${\cal{H}}= \oplus _{M=0}^N{\cal{H}}^{(M)}$
according to the action of the operator $ {\cal{Q}}^{(1)}$ given by
(\ref{Q-operator}): \bea {\cal{Q}}^{(1)}\,{\cal{H}}^{(M)}=M
\,{\cal{H}}^{(M)},\quad M=0,1,\cdots,N. \eea The commutativity of
the transfer matrices and the operator $ {\cal{Q}}^{(1)}$  implies
that each of the subspace is invariant under $t_m(u)$. Hence the
whole set of eigenvalues of the transfer matrices can be decomposed
into $N+1$ series. Acting the operators in (\ref{Id-2})-(\ref{Id-5})
on any subspace ${\cal{H}}^{(M)}$, we obtain the asymptotic
behaviors of the corresponding $\Lambda_m(u)$ \bea
\Lambda_1(u)|_{u\rightarrow +\infty}&=&-
\frac{e^{(2N+4)u+3\eta}}{4^{N+1}}\left[c\frac{c_1'c_2'}{c'}e^{2M\eta+\eta}+(c_1c_2'+c_1'c_2e^{2\eta})e^{(N-M)\eta}\right]+\cdots, \label{Ld-2}\\[4pt]
\Lambda_1(u)|_{u\rightarrow -\infty}&=&-
\frac{e^{-(2N+4)u-3\eta}}{4^{N+1}}\left[c'\frac{c_1c_2}{c}e^{-2M\eta+\eta}
+(c_1c_2'+c_1'c_2e^{2\eta})e^{(M-N)\eta}\right]+\cdots, \label{Ld-3}\\[4pt]
\Lambda_2(u)|_{u\rightarrow +\infty}&=&-
\frac{1}{4^{2N+3}}e^{4(N+3)u+4\eta}c\frac{c_1'c_2'}{c'}\left[c'\frac{c_1c_2}{c}e^{-2M\eta+\eta}\right.\no\\[4pt]
&&\left.+(c_1c_2'+c_1'c_2e^{2\eta})e^{(M-N)\eta}\right]+\cdots, \label{Ld-4}\\[4pt]
\Lambda_2(u)|_{u\rightarrow -\infty}&=&-
\frac{1}{4^{2N+3}}e^{-4(N+3)u-2\eta}c'\frac{c_1c_2}{c}\left[c\frac{c_1'c_2'}{c'}e^{2M\eta+\eta}\right.\no\\[4pt]
&&\left.+(c_1c_2'+c_1'c_2e^{2\eta})e^{(N-M)\eta}\right]+\cdots.
\label{Ld-5} \eea

Therefore, the functional relations (\ref{Ld-1-0})-(\ref{Ld-1-3}), the values at
special points (\ref{LLd-1})-(\ref{LLd-2}) and the asymptotic
behaviors\footnote{It is remarked that for elliptical integrable models  asymptotic behaviors such as (\ref{Ld-2})-(\ref{Ld-5}) will be
replaced by the associated quasi-periodicities of the fused transfer matrices (for an example, see \cite{Cao14NB886} (or \cite{Cao13NB877}) for the XYZ closed chain (or the XYZ open chain)).} (\ref{Ld-2})-(\ref{Ld-5})  can
provide us sufficient conditions to completely determine the
corresponding eigenvalues $\Lambda_m(u)$.

\section{Nested inhomogeneous $T-Q$ relation}
\label{T-Q} \setcounter{equation}{0}

Now we construct the eigenvalues $\Lambda_m(u)$ of the fused
transfer matrices $t_m(u)$. For simplicity, we define some functions
\begin{eqnarray}
&& b_0(u)=\prod_{j=1}^N\sinh(u-\theta_j)\sinh(u+\theta_j), \quad
a_0(u)=b_0(u+\eta), \label{10-14-3-1118} \\
&&Q^{(k)}(u)=\prod_{l=1}^{L_k}\sinh(u-\lambda^{(k)}_l)\sinh(u+\lambda^{(k)}_l+
k \eta), \quad k=1,2, \label{Q-functions-Open-1}
\end{eqnarray}
where $L_1$ and $L_2$ are non-negative integers. Due to the survived
$U(1)$ conserved charge ${\cal{Q}}^{(1)}$ in the system, the number
of one kind of Bethe roots can be chosen as $M$, which is similar as
the algebraic Bethe ansatz. Without losing generality, we put
$L_2=M$. In order to construct the eigenvalues of the fused transfer
matrices, we introduce three $\tilde z(u)$ functions
\begin{eqnarray}
&&\tilde z_{1}(u)=z_{1}(u)+x_{1}(u), \quad \tilde
z_{2}(u)=z_{2}(u), \quad \tilde z_{3}(u)=z_{3}(u).
\end{eqnarray}
Here $z_{m}(u)$ is defined as
\begin{eqnarray}
&&z_{m}(u) =\frac{\sinh (2u)\sinh(2u+3
\eta)}{\sinh(2u+{(m-1)}\eta)\sinh(2u+ {m}\eta)}K^{(m)}(u)b_0(u)
\frac{Q^{(m-1)}(u+\eta)Q^{(m)}(u-\eta)}{Q^{(m-1)}(u)Q^{(m)}(u)},\no\\[6pt]
&& \hspace{5cm} m=1,2,3, \label{2z2-1118-1216}
\end{eqnarray}
with the notations $Q^{(0)}(u)=b_0(u)$, $Q^{(3)}(u)=1$ and
$x_1(u)$ is defined as
\begin{eqnarray}
x_{1}(u)= \sinh (2u)\sinh(2u+  3 \eta)a_0(u)b_0(u)
\frac{f_1(u)Q^{(2)}(-u-\eta)}{Q^{(1)}(u)},
\end{eqnarray}
where $K^{(m)}(u)$ are the decompositions of the quantum determinant
and $f_1(u)$ is a function which will be determined later.

The nested functional $T-Q$ ansatz is expressed as
\begin{eqnarray}
\Lambda(u)&=&\sum_{i_1=1}^{3} \tilde z_{i_1}(u) \no \\[4pt]
&=&\frac{\sinh(2u+3\eta)}{\sinh(2u+\eta)} K^{(1)}(u)a_0(u)
\frac{Q^{(1)}(u-\eta)}{Q^{(1)}(u)}
\nonumber \\[4pt]
&&\,+ \frac{\sinh
(2u)\sinh(2u+3\eta)}{\sinh(2u+\eta)\sinh(2u+2\eta)}K^{(2)}(u)b_0(u)\frac{Q^{(1)}(u+\eta)Q^{(2)}(u-\eta)}{Q^{(1)}(u)Q^{(2)}(u)}
\nonumber \\[4pt]
&&\, + \frac{\sinh (2u)}{\sinh(2u+2\eta)} K^{(3)}(u)b_0(u)
\frac{Q^{(2)}(u+\eta)}{Q^{(2)}(u)}\no\\[4pt]
&&\, +\,\sinh (2u)\sinh(2u+3\eta)a_0(u)b_0(u)
\frac{f_1(u)Q^{(2)}(-u-\eta)}{Q^{(1)}(u)}, \label{3t1-101}
\\[6pt]
\Lambda_2(u)&=&\rho_2(2u-\eta)\left[\sum_{1\leq i_1<i_2 \leq 3 }
\tilde z_{i_1}(u) \tilde z_{i_2}(u-\eta) - x_1(u) z_2
(u-\eta)\right] \no \\[4pt]
&=&\rho_2(2u-\eta)b_0(u-\eta)\no\\[4pt]
&&\times\left\{ \frac{\sinh(2u-2\eta)\sinh(2u+3\eta)}{\sinh
(2u)\sinh(2u-\eta)} K^{(1)}(u)K^{(2)}(u-\eta)a_0(u)
\frac{Q^{(2)}(u-2\eta)}{Q^{(2)}(u-\eta)}
\right.\nonumber \\[4pt]
&& \, \left. +
\frac{\sinh(2u-2\eta)\sinh(2u+3\eta)}{\sinh(2u+\eta)\sinh(2u)}K^{(1)}(u)K^{(3)}(u-\eta)a_0(u)\frac{Q^{(1)}(u-\eta)Q^{(2)}(u)}{Q^{(1)}(u)Q^{(2)}(u-\eta)}
\right.\nonumber \\[4pt]
&&\, +
\frac{\sinh(2u-2\eta)\sinh(2u+3\eta)}{\sinh(2u+\eta)\sinh(2u+2\eta)}
K^{(2)}(u)K^{(3)}(u-\eta)b_0(u)
\frac{Q^{(1)}(u+\eta)}{Q^{(1)}(u)}\no \\[4pt]
&& \, +\left.\frac{\sinh (2u-2\eta)}{\sinh (2u)}
\,\sinh (2u) \sinh(2u+{3}\eta)a_0(u)b_0(u)f_1(u)\frac{Q^{(2)}(-u-\eta)}{Q^{(1)}(u)}\right.\no\\[4pt]
&& \left. \qquad \times \frac{Q^{(2)}(u)K^{(3)}(u-\eta)}{
Q^{(2)}(u-\eta)}\right\}, \label{3t2-101}
\\[4pt]
\Lambda_3(u)&=&\prod_{k=1}^3\rho_2(2u-k\eta)
z_{1}(u)z_{2}(u-\eta)z_{3}(u-2\eta),\label{t2-3-3}
\end{eqnarray}
where the non-negative integer $L_1$ is \bea L_1=N+M+6.\label{LN-1}
\eea

Because the eigenvalues $\Lambda_m(u)$ are the trigonometric
polynomials, the residues of right hand sides of
Eqs.(\ref{3t1-101})-(\ref{t2-3-3}) should be zero, which gives the
constraints of the Bethe roots $\{\lambda^{(r)}_l\}$ thus the Bethe
ansatz equations. The Bethe ansatz equations obtained from the
regularity of $\Lambda(u)$ should be the same as that obtained from
the regularity of $\Lambda_2(u)$. The function $Q^{(r)}(u)$ has two
zero points, $\lambda^{(r)}_l$ and $-\lambda^{(r)}_l-r\eta$. The
Bethe ansatz equations obtained from these two points also should be
the same, which requires
\begin{eqnarray}K^{(r)}(u) K^{(r)}(-u-r\eta)=K^{(r+1)}(u) K^{(r+1)}(-u-r\eta),
\quad r=1, 2. \label{klkl} \end{eqnarray} We note that
$\Lambda_3(u)$ is a trigonometric polynomial automatically. The fact
that $\Lambda_3(u)$ should be the quantum determinant requires \bea
K^{(1)}(u)K^{(2)}(u-\eta)K^{(3)}(u-2\eta)=\frac{\Delta_q\{K^-(u)\}\Delta_q\{K^+(u)\}}{\prod_{k=1}^3\sinh(2u+k\eta)\sinh(2u-(k+1)\eta)}.
\eea The consistency of Bethe ansatz equations also require that the
function $f_1(u)$ has the crossing symmetry
\begin{eqnarray}
f_1(u) = f_1(-u-\eta). \label{f11-103-1-1106-1216}
\end{eqnarray}

Furthermore, the eigenvalues $\Lambda_m(u)$ should satisfy the
functional relations (\ref{Ld-1}). This gives other constraints of
the function $f_1(u)$. Considering all the above requirements, we
parameterize the function $f_1(u)$ as
\begin{eqnarray}
f_1(u)=  (\frac 1 4)^3h \sinh (2u)\sinh^2(2u+\eta)
\sinh(2u+2\eta)\sinh(2u-\eta)\sinh(2u+{3}\eta),
\end{eqnarray}
where $h$ is a constant which is determined by the asymptotic
behaviors of the $\Lambda_m(u)$.

Now, we are ready to give the Bethe ansatz equations as following
\bea &&1+\frac{\sinh
(2\lambda_l^{(1)})}{\sinh(2\lambda_l^{(1)}+2\eta)}\frac{K^{(2)}({\lambda_l^{(1)}})b_0(\lambda^{(1)}_l)}{K^{(1)}({\lambda_l^{(1)}})a_0(\lambda^{(1)}_l)}
\frac{Q^{(1)}(\lambda^{(1)}_l+\eta)Q^{(2)}(\lambda^{(1)}_l-\eta)}{Q^{(1)}(\lambda^{(1)}_l-\eta)Q^{(2)}(\lambda^{(1)}_l)}\no\\[6pt]
&&\quad\quad\quad=-h\,{\sinh^2(2\lambda^{(1)}_l)\sinh^3(2\lambda^{(1)}_l+\eta)\sinh(2\lambda^{(1)}_l+2\eta)\sinh(2\lambda^{(1)}_l+3\eta)
}\no\\[6pt]
&&\quad \quad \quad\quad \times \sinh(2\lambda^{(1)}_l-\eta)
\frac{b_0(\lambda^{(1)}_l)Q^{(2)}(\lambda^{(1)}_l-\eta)}
{4^3K^{(1)}(\lambda^{(1)}_l)Q^{(1)}(\lambda^{(1)}_l-\eta)},\,\,\,\no\\[4pt]
&&\quad\quad l=1,\ldots, M+N+6,\label{BAE-Open-3-1} \\[6pt]
&&\frac{\sinh(2\lambda_k^{(2)}+{3}\eta)}{\sinh(2\lambda_k^{(2)}+\eta)}\frac{K^{(2)}({\lambda_k^{(2)}})}{K^{(3)}({\lambda_k^{(2)}})}
\frac{Q^{(1)}(\lambda^{(2)}_k+\eta)Q^{(2)}(\lambda^{(2)}_k-\eta)}{Q^{(1)}(\lambda^{(2)}_k)Q^{(2)}(\lambda^{(2)}_k+\eta)}=-1,
 \,\, k=1,\ldots, M.\label{BAE-Open-3-2} \eea
It is easy to check that the Bethe ansatz equations
(\ref{BAE-Open-3-1})-(\ref{BAE-Open-3-2}) guarantee the regularities
of the ansatz $\Lambda(u)$ given by (\ref{3t1-101}) and the ansatz
$\Lambda_2(u)$ given by (\ref{3t2-101}). Moreover, the ansatz
(\ref{3t1-101})-(\ref{t2-3-3}) indeed satisfy the function relations
(\ref{Ld-1}).

The left tasks are to determine the value of $h$ in the ansatz
(\ref{3t1-101})-(\ref{t2-3-3}), which can be done by analyzing the
asymptotic behaviors, and to check the consistency of values at the
special points (\ref{LLd-1})-(\ref{LLd-2}). Because there are three
kinds of boundary reflection matrices and they give the different
behaviors, let us consider them one by one.

For the reflection matrices $K^{\pm}$ given by (\ref{K-matrix-1})
and (\ref{K-matrix-4}), the decomposition $K^{(i)}(u)$ can be chosen
as \bea &&K^{(1)}(u)=e^{2\eta}(e^{-u}\sinh(\zeta+u)-ce^{-2u}\sinh
2u)
\no\\[4pt]
&&\hspace{2cm}\times (e^{u-\frac {\eta}
2}\sinh(\zeta'-u+\frac{\eta}{2})+c'e^{2(u-\frac {\eta}
2)}\sinh 2(u-\frac {\eta} 2)),\label{kkk1-0} \\[4pt]
&&K^{(2)}(u)=e^{2\eta}(e^{u+\eta}\sinh(\zeta-u-\eta)+ce^{2(u+\eta)}\sinh
2(u+\eta))\no\\[4pt]
&&\hspace{2cm}\times(e^{-u-\frac {3\eta}
2}\sinh(\zeta'+u+\frac{3\eta}{2})-c'e^{-2(u+\frac {3\eta} 2)}\sinh
2(u+\frac {3\eta} 2)),\label{kkk1-1} \\[4pt]
&&K^{(3)}(u)= K^{(2)}(u). \label{kkk1}\eea From the asymptotic
behaviors (\ref{Ld-2})-(\ref{Ld-5}) of the corresponding
trigonometric polynomials, we arrive at the value of $h$
\begin{eqnarray}
&&h=c\frac{c_1'c_2'}{c'}e^{(M+N)\eta+15\eta}+c'\frac{c_1c_2}{c}e^{-(M+N)\eta-13\eta}
-(c_1c_2'+c_1'c_2e^{2\eta}).\label{r3-1}
\end{eqnarray}
Then one can check that the values of ansatz
(\ref{3t1-101})-(\ref{3t2-101}) at the special points
(\ref{LLd-1})-(\ref{LLd-2}) are the same as those of the
corresponding fused transfer matrices, and we finish our
construction.

Taking the homogeneous limit $\{\theta_j=0, j=1, \ldots, N\}$, we
conclude that the $T-Q$ relation $\Lambda(u)$ given by
(\ref{3t1-101}) is the eigenvalue of the transfer matrices $t(u)$ of
the trigonometric $SU(3)$ open spin chain with the most general
off-diagonal integrable boundary conditions. The energy of the
Hamiltonian (\ref{hamimi}) reads
\begin{eqnarray} &&E=\sinh\eta
\frac{\partial \ln \Lambda(u)}{\partial u}|_{u=0,\{\theta_j\}=0},\label{energy_E}
\end{eqnarray}
where the Bethe roots should satisfy the Bethe ansatz equations
(\ref{BAE-Open-3-1})-(\ref{BAE-Open-3-2}). Above results can be
reduced to the diagonal boundaries ones obtained by the algebraic
Bethe ansatz \cite{8Mel054,8Mel055,8Mel056}.

\begin{figure}[!htbp]
\centering
\includegraphics[height=75mm,width=100mm]{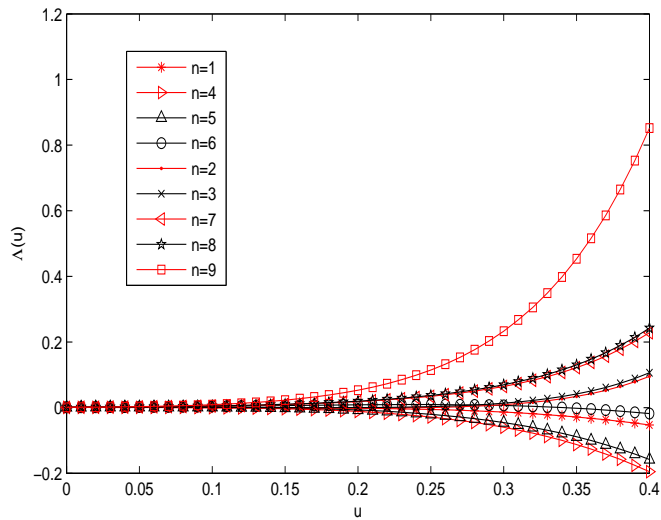}
\caption{(color online) $\Lambda(u)$ vs. $u$ for the case of  $N=2$ and $\{\theta_j=0\}$. The curves calculated from $T-Q$ relation (\ref{3t1-101}) and the nested BAEs (\ref{BAE-Open-3-1})-(\ref{BAE-Open-3-2}) are exactly the same as those obtained from the exact diagonalization of the transfer matrix $t(u)$.}\label{Lambda_u}
\end{figure}

Numerical solutions of the BAEs
(\ref{BAE-Open-3-1})-(\ref{BAE-Open-3-2}) for small size\footnote{It
is still an interesting open problem to investigate the root pattern
of the BAEs (\ref{BAE-Open-3-1})-(\ref{BAE-Open-3-2}) associated
with  the inhomogeneous  $T-Q$ relation. Nevertheless, a standard
method to study the thermodynamic limit was developed  in
\cite{Li14} by considering a sequence of discrete $\eta$ values at
which the inhomogeneous terms in the BAEs vanish and in the
thermodynamic limit these discrete $\eta$ values become dense.} with
a random choice of $\eta$ imply that the BAEs indeed  give the
complete solutions of the model. Here we present the result for the
$N=2$ case: the numerical solutions  of the BAEs for the $N=2$ case
are shown in Table \ref{tab:BAEs}, while the calculated $\Lambda(u)$
curves for the case of $N=2$ are shown in Figure \ref{Lambda_u}.

\begin{sidewaystable}[htbp]
\caption{Solutions of BAEs (\ref{BAE-Open-3-1})-(\ref{BAE-Open-3-2}) for the case of $N=2, \eta=0.3, \{\theta_j=0\}, \zeta=0.1, c=1.0, c_{1}=-0.5, \zeta'=-0.1, c'=-0.5,c_{1}'=-0.7$. The symbol $n$ indicates the number of the eigenvalues, and $E_{n}$ is the corresponding eigenenergy. The energy $E_{n}$ calculated from (\ref{energy_E}) is the same as that from the exact diagonalization of the Hamiltonian (\ref{hamimi}).}
\centering 
\begin{footnotesize}
\begin{tabular}{|c|c|c|c|c|c|c|c|c|c|c|c|c|c|c|c|c|c|}
\hline\hline $\lambda_1^{(1)}$ & $\lambda_2^{(1)}$ & $\lambda_3^{(1)}$ & $\lambda_4^{(1)}$ & $\lambda_5^{(1)}$ & $\lambda_6^{(1)}$ & $\lambda_7^{(1)}$ \\ \hline
$0.4091+0.1448i$  &  $0.4091-0.1448i$  &  $0.4942+1.3424i$  &  $0.4942-1.3424i$  &  $0.2366-0.0000i$  &  $-0.1500-0.1466i$  &  $-0.7730+0.0000i$  \\
 $0.3823+0.1863i$  &  $0.3823-0.1863i$  &  $0.4842+1.3499i$  &  $0.4842-1.3499i$  &  $0.1292-0.0000i$  &  $-0.1500+0.3054i$  &  $0.3917-0.0000i$  \\
 $-0.7839-1.3503i$  &  $-0.7839+1.3503i$  &  $-0.6860+0.1856i$  &  $-0.6860-0.1856i$  &  $0.0983-0.0000i$  &  $-0.1500-0.2984i$  &  $0.3884+0.0000i$  \\
 $0.3461+0.1125i$  &  $0.3461-0.1125i$  &  $0.4989+1.3364i$  &  $0.4989-1.3364i$  &  $0.3242-0.0000i$  &  $0.1351+0.0000i$  &  $0.4436-0.0000i$  \\
 $0.4981-1.3375i$  &  $0.4981+1.3375i$  &  $0.3679+0.1168i$  &  $0.3679-0.1168i$  &  $0.2843-0.0000i$  &  $0.5641-1.5708i$  &  $0.4502-0.0000i$  \\
 $0.0946+0.0271i$  &  $0.0946-0.0271i$  &  $0.4139+0.1390i$  &  $0.4139-0.1390i$  &  $0.4947+1.3419i$  &  $0.4947-1.3419i$  &  $0.4722-0.0000i$  \\
 $0.1029+0.0175i$  &  $0.1029-0.0175i$  &  $0.4165+0.1626i$  &  $0.4165-0.1626i$  &  $0.4870+1.3478i$  &  $0.4870-1.3478i$  &  $0.4822+0.0000i$  \\
 $0.4211+0.1625i$  &  $0.4211-0.1625i$  &  $0.0997+0.0211i$  &  $0.0997-0.0211i$  &  $0.4868+1.3482i$  &  $0.4868-1.3482i$  &  $0.4855-0.0000i$  \\
 $-0.3852-0.0302i$  &  $-0.3852+0.0302i$  &  $0.4432+0.1924i$  &  $0.4432-0.1924i$  &  $0.5110+0.0377i$  &  $0.5110-0.0377i$  &  $0.4786+1.3557i$  \\
\hline\hline $\lambda_8^{(1)}$ & $\lambda_9^{(1)}$ & $\lambda_{10}^{(1)}$ & $\lambda_1^{(2)}$ & $\lambda_2^{(2)}$  & $E_{n}$ & $n$  \\ \hline
 $0.5592-1.5708i$ &--&--&--&--& $-4.932732$ & $1$  \\
 $0.4614+0.0000i$  &  $0.5501-1.5708i$  &--&  $-0.3000+0.2629i$ &--& $0.044922$ & $2$  \\
 $-0.7641+0.0000i$  &  $-0.8497+1.5708i$  &--&  $-0.3000-0.2664i$ &--& $1.356985$ & $3$  \\
 $0.5651-1.5708i$ &--&--&--&--& $2.206441$ & $4$  \\
 $0.1007+0.0000i$ &--&--&--&--& $3.696554$ & $5$  \\
 $0.5597+1.5708i$ &--&--&--&--& $6.612138$ & $6$  \\
 $0.3507-0.0000i$  &  $-0.8526-1.5708i$  &--&  $0.0104+0.0000i$ &--& $6.923080$ & $7$  \\
 $0.3483-0.0000i$  &  $-0.8522+1.5708i$  &--&  $0.0235-0.0000i$ &--& $7.079096$ & $8$  \\
 $0.4786-1.3557i$  &  $0.2226+0.0000i$  &  $-0.8440-1.5708i$  &  $-0.0527+0.0000i$  &  $0.1841+0.0000i$ & $8.942457$ & $9$  \\
 \hline\hline\end{tabular} \label{tab:BAEs}
 \end{footnotesize}
\end{sidewaystable}

\section{Diagonal boundary case}
\label{diag} \setcounter{equation}{0}

When the parameters $c,c_1,c_2,c',c_1',c_2'$ in the reflection matrices
$K^{\pm}(u)$ given by (\ref{K-matrix-1}) and (\ref{K-matrix-4}) vanish,
the corresponding $K$-matrices become diagonal ones. Let us denote them by $\bar{K}^{\pm}(u)$, namely,
\bea
&&\bar{K}^-(u)=\left(\begin{array}{ccc}e^{u}\sinh(\zeta-u)&0&0\\[6pt]
0&e^{u}\sinh(\zeta-u)&\\[6pt]
0&&e^{-u}\sinh(\zeta+u)\end{array}\right),
\label{K-matrix-D-1}\\[6pt]
&&\bar{K}^+(u)={\cal
M}\bar{K}^-(-u-3\eta/2)\left|_{\zeta\rightarrow\zeta'}\right..\label{K-matrix-D-2} \eea
Moreover, the corresponding decomposition $\{K^{(i)}(u)\}$ in (\ref{kkk1-0})-(\ref{kkk1}), if denoted by
$\bar{K}^{(i)}(u)$, are given by
\bea
&&\bar{K}^{(1)}(u)=e^{\frac{3}{2}\eta}\sinh(\zeta+u)\sinh(\zeta'-u+\frac{\eta}{2}),\label{kkk1-D-0} \\[4pt]
&&\bar{K}^{(2)}(u)=e^{\frac{3}{2}\eta}\sinh(\zeta-u-\eta)\sinh(\zeta'+u+\frac{3\eta}{2}),\label{kkk1-D-1} \\[4pt]
&&\bar{K}^{(3)}(u)= \bar{K}^{(2)}(u). \label{kkk1-D}\eea Then the corresponding $T-Q$ relations (\ref{3t1-101})-(\ref{3t2-101}) become the usual
homogeneous  ones and now are given by
\begin{eqnarray}
\Lambda(u)&=&
\frac{\sinh(2u+3\eta)}{\sinh(2u+\eta)} \bar{K}^{(1)}(u)a_0(u)
\frac{\bar{Q}^{(1)}(u-\eta)}{\bar{Q}^{(1)}(u)}
\nonumber \\[4pt]
&&\,+ \frac{\sinh
2u\sinh(2u+3\eta)}{\sinh(2u+\eta)\sinh(2u+2\eta)}\bar{K}^{(2)}(u)b_0(u)\frac{\bar{Q}^{(1)}(u+\eta)\bar{Q}^{(2)}(u-\eta)}{\bar{Q}^{(1)}(u)\bar{Q}^{(2)}(u)}
\nonumber \\[4pt]
&&\, + \frac{\sinh 2u}{\sinh(2u+2\eta)} \bar{K}^{(3)}(u)b_0(u)
\frac{\bar{Q}^{(2)}(u+\eta)}{\bar{Q}^{(2)}(u)}, \label{T-Q-D-1}
\\[6pt]
\Lambda_2(u)
&=&\rho_2(2u-\eta)b_0(u-\eta)\no\\[4pt]
&&\times\left\{ \frac{\sinh(2u-2\eta)\sinh(2u+3\eta)}{\sinh
2u\sinh(2u-\eta)} \bar{K}^{(1)}(u)\bar{K}^{(2)}(u-\eta)a_0(u)
\frac{\bar{Q}^{(2)}(u-2\eta)}{\bar{Q}^{(2)}(u-\eta)}
\right.\nonumber \\[4pt]
&& \, \left. +
\frac{\sinh(2u-2\eta)\sinh(2u+3\eta)}{\sinh(2u+\eta)\sinh2u}\bar{K}^{(1)}(u)\bar{K}^{(3)}(u-\eta)a_0(u)
\frac{\bar{Q}^{(1)}(u-\eta)\bar{Q}^{(2)}(u)}{\bar{Q}^{(1)}(u)\bar{Q}^{(2)}(u-\eta)}
\right.\nonumber \\[4pt]
&&\, +\left.
\frac{\sinh(2u-2\eta)\sinh(2u+3\eta)}{\sinh(2u+\eta)\sinh(2u+2\eta)}
\bar{K}^{(2)}(u)\bar{K}^{(3)}(u-\eta)b_0(u)
\frac{\bar{Q}^{(1)}(u+\eta)}{\bar{Q}^{(1)}(u)}\rt\}. \label{T-Q-D-2}
\end{eqnarray} Here the corresponding $Q$-functions are \bea \bar{Q}^{(k)}(u)
=\prod_{l=1}^{\bar{M}_k}\sinh(u-\lambda^{(k)}_l)\sinh(u+\lambda^{(k)}_l+
k \eta),\quad 0\leq\bar{M}_2\leq N,\quad 0\leq \bar{M}_1\leq
\bar{M}_2. \eea The resulting homogeneous relation (\ref{T-Q-D-1})
recovers that obtained by the algebraic Bethe ansatz method
\cite{8Mel056}, while the reference state is chosen as \bea
|vac>=\left(\begin{array}{c}0\\0\\1
\end{array}\right)\otimes \left(\begin{array}{c}0\\0\\1
\end{array}\right) \otimes \left(\begin{array}{c}0\\0\\1
\end{array}\right)\otimes\cdots \otimes \left(\begin{array}{c}0\\0\\1
\end{array}\right), \label{vac-1}\eea
and the associated creation operators are the off-diagonal matrix elements of the 3-rd row of
the double-row monodromy matrix $\mathbb{T}(u)$ given by (\ref{Mon-V-0}).

\section{Conclusions}

In this paper, we study the exact solution of the anisotropic
quantum spin chain with generic open boundary conditions and
associated with $SU_q(3)$ algebra. After giving the off-diagonal
reflection matrixes, by using the fusion technique, we obtain some
closed operator identities among the transfer matrices, the
degenerate points and the corresponding asymptotic behaviors. Based
on them, we construct the nested inhomogeneous $T-Q$ relations and
the Bethe ansatz equations. These results can be generalized to the
higher rank case. Moreover, when the boundary parameters take
special values corresponding to the diagonal reflection matrices,
our results recover those previously obtained by the conventional
Bethe ansatzs.

\section*{Acknowledgments}

We would like to thank Y. Wang for his valuable discussions and continuous  encouragement.
The financial supports from the National Natural Science Foundation
of China (Grant Nos. 11375141, 11374334, 11434013, 11425522 and
11547045), BCMIIS and the Strategic Priority Research Program of the
Chinese Academy of Sciences are gratefully acknowledged.

\section*{Appendix: Asymptotic
behaviors for the other two cases reflecting matrices}
\setcounter{equation}{0}
\renewcommand{\theequation}{A.\arabic{equation}}

When the reflection matrices $K^{\pm}$ are given by
(\ref{K-matrix-2}) and (\ref{K-matrix-4}), the asymptotic
behaviors of fused transfer matrices $t_m(u)$ are
 \bea t_1(u)|_{u\rightarrow +\infty}&=&-
\frac{1}{4^{N+1}}e^{(2N+4)u+3\eta}\left[c'\frac{c_1c_2}{c}e^{2\eta}\,e^{2\eta {\cal{Q}}^{(2)}}\rt.\no\\[4pt]
&&+\left.(c_1c_2'+c_1'c_2e^{4\eta})e^{N\eta}e^{-\eta {\cal{Q}}^{(2)}}\right]+\cdots, \label{aId-2}\\[4pt]
t_1(u)|_{u\rightarrow -\infty}&=&-
\frac{1}{4^{N+1}}e^{-(2N+4)u-3\eta}\left[c\frac{c_1'c_2'}{c'}e^{2\eta}\,e^{-2\eta {\cal{Q}}^{(2)}}\right.\no\\[4pt]
&&\left.+(c_1c_2'+c_1'c_2e^{4\eta})\,e^{-N\eta}\,e^{\eta {\cal{Q}}^{(2)}}\right]+\cdots, \label{aId-3}\\[4pt]
t_2(u)|_{u\rightarrow +\infty}&=&-
\frac{1}{4^{2N+3}}e^{4(N+3)u+5\eta}c'\frac{c_1c_2}{c}\left[c\frac{c_1'c_2'}{c'}e^{2\eta}\, e^{-2\eta {\cal{Q}}^{(2)}}\right.\no\\[4pt]
&&\left.+(c_1c_2'+c_1'c_2e^{4\eta})e^{-N\eta}\,e^{\eta {\cal{Q}}^{(2)}}\right]+\cdots, \label{aId-4}\\[4pt]
t_2(u)|_{u\rightarrow -\infty}&=&-
\frac{1}{4^{2N+3}}e^{-4(N+3)u-\eta}c\frac{c_1'c_2'}{c'}\left[c'\frac{c_1c_2}{c}e^{2\eta}e^{2\eta {\cal{Q}}^{(2)}}\right.\no\\[4pt]
&&\left.+(c_1c_2'+c_1'c_2e^{4\eta})\,e^{N\eta}\,e^{-\eta {\cal{Q}}^{(2)}}\right]+\cdots,
\label{aId-5} \eea
where the operator ${\cal{Q}}^{(2)}$ is
\bea
{\cal{Q}}^{(2)}=\sum_{l=1}^N E^{22}_l,\quad E^{22}=\left(\begin{array}{ccc}0&0&0\\0&1&0\\0&0&0\end{array}\rt). \label{Q-operator-2}
\eea

The asymptotic behaviors of fused transfer matrices $t_m(u)$
associated with the reflection matrices $K^{\pm}$ given by
(\ref{K-matrix-3}) and (\ref{K-matrix-4}) are
 \bea t_1(u)|_{u\rightarrow +\infty}&=&-
\frac{1}{4^{N+1}}e^{(2N+4)u+5\eta}\left[c'\frac{c_1c_2}{c}e^{\eta}e^{2\eta {\cal{Q}}^{(3)}}\right.\no\\[4pt]
&&\left.+(c_1c_2'+c_1'c_2e^{2\eta})\,e^{N\eta}\,e^{-\eta {\cal{Q}}^{(3)}}\right]+\cdots, \label{aId-6}\\[4pt]
t_1(u)|_{u\rightarrow -\infty}&=&-
\frac{1}{4^{N+1}}e^{-(2N+4)u-\eta}\left[c\frac{c_1'c_2'}{c'}e^{\eta}e^{-2\eta {\cal{Q}}^{(3)}}\right.\no\\[4pt]
&&\left.+(c_1c_2'+c_1'c_2e^{2\eta})\,e^{-N\eta}\,e^{\eta {\cal{Q}}^{(3)}}\right]+\cdots, \label{aId-7}\\[4pt]
t_2(u)|_{u\rightarrow +\infty}&=&-
\frac{1}{4^{2N+3}}e^{4(N+3)u+8\eta}c'\frac{c_1c_2}{c}\left[c\frac{c_1'c_2'}{c'}e^{\eta}\,e^{-2\eta {\cal{Q}}^{(3)}}\right.\no\\[4pt]
&&\left.+(c_1c_2'+c_1'c_2e^{2\eta})\,e^{-N\eta}\,e^{\eta {\cal{Q}}^{(3)}}\right]+\cdots, \label{aId-8}\\[4pt]
t_2(u)|_{u\rightarrow -\infty}&=&-
\frac{1}{4^{2N+3}}e^{-4(N+3)u+2\eta}c\frac{c_1'c_2'}{c'}\left[c'\frac{c_1c_2}{c}e^{\eta}\,e^{2\eta {\cal{Q}}^{(3)}}\right.\no\\[4pt]
&&\left.+(c_1c_2'+c_1'c_2e^{2\eta})\,e^{N\eta}\, e^{-\eta {\cal{Q}}^{(3)}}\right]+\cdots,
\label{aId-9} \eea
where the operator ${\cal{Q}}^{(3)}$ is
\bea
{\cal{Q}}^{(3)}=\sum_{l=1}^N E^{33}_l,\quad E^{33}=\left(\begin{array}{ccc}0&0&0\\0&0&0\\0&0&1\end{array}\rt). \label{Q-operator-3}
\eea

Some remarks are in order. Similarly as the operator ${\cal{Q}}^{(1)}$ given by (\ref{Q-operator}), the above operator ${\cal{Q}}^{(2)}$ (resp.
${\cal{Q}}^{(3)}$) generates a remaining $U(1)$-symmetry for the corresponding model respectively. Moreover the eigenvalues of the operator
are $0,\,1,\cdots,N$. Hence the whole Hilbert space can be decomposed into the subspaces labeled by its eigenvalue, on which the transfer matrices
are invariant. We can calculate the asymptotic behaviors of the corresponding transfer matrices on each subspace.

For the reflection matrix $K^{\pm}$ given by (\ref{K-matrix-2})
and (\ref{K-matrix-4}), the asymptotic behaviors of $\Lambda_m(u)$
can be obtained by acting the operator (\ref{aId-2})-(\ref{aId-5})
on the subspace on which the eigenvalue of the operator ${\cal{Q}}^{(2)}$ is $M$.
After some calculations, we arrive at
\bea \Lambda_1(u)|_{u\rightarrow +\infty}&=&-
\frac{e^{(2N+4)u+3\eta}}{4^{N+1}}\left[c'\frac{c_1c_2}{c}e^{2M\eta+2\eta}+(c_1c_2'+c_1'c_2e^{4\eta})e^{(N-M)\eta}\right]+\cdots, \label{k2Ld-2}\\[4pt]
\Lambda_1(u)|_{u\rightarrow -\infty}&=&-
\frac{1}{4^{N+1}}e^{-(2N+4)u-3\eta}\left[c\frac{c_1'c_2'}{c'}e^{-2M\eta+2\eta}
\right.\no\\[4pt]
&&\left.+(c_1c_2'+c_1'c_2e^{4\eta})e^{(M-N)\eta}\right]+\cdots, \label{k2Ld-3}\\[4pt]
\Lambda_2(u)|_{u\rightarrow +\infty}&=&-
\frac{1}{4^{2N+3}}e^{4(N+3)u+5\eta}c'\frac{c_1c_2}{c}\left[c\frac{c_1'c_2'}{c'}e^{-2M\eta+2\eta}
\right.\no\\[4pt]
&&\left.+(c_1c_2'+c_1'c_2e^{4\eta})e^{(M-N)\eta}\right]+\cdots, \label{k2Ld-4}\\[4pt]
\Lambda_2(u)|_{u\rightarrow -\infty}&=&-
\frac{1}{4^{2N+3}}e^{-4(N+3)u-\eta}c\frac{c_1'c_2'}{c'}\left[c'\frac{c_1c_2}{c}e^{2M\eta+2\eta}\right.\no\\[4pt]
&&\left.+(c_1c_2'+c_1'c_2e^{4\eta})e^{(N-M)\eta}\right]+\cdots.
\label{k2Ld-5}\eea

For the reflection matrix $K^{\pm}$ given by (\ref{K-matrix-3})
and (\ref{K-matrix-4}), the asymptotic behaviors of $\Lambda_m(u)$
can be obtained by acting the operator (\ref{aId-6})-(\ref{aId-9})
on the subspace on which the eigenvalue of the operator ${\cal{Q}}^{(3)}$ is $M$.
The finial results read \bea
\Lambda_1(u)|_{u\rightarrow +\infty}&=&-
\frac{e^{(2N+4)u+5\eta}}{4^{N+1}}\left[c'\frac{c_1c_2}{c}e^{2M\eta+\eta}+(c_1c_2'+c_1'c_2e^{2\eta})e^{(N-M)\eta}\right]+\cdots, \label{k3Ld-2}\\[4pt]
\Lambda_1(u)|_{u\rightarrow -\infty}&=&-
\frac{1}{4^{N+1}}e^{-(2N+4)u-\eta}\left[c\frac{c_1'c_2'}{c'}e^{-2M\eta+\eta}
\right.\no\\[4pt]
&&\left.+(c_1c_2'+c_1'c_2e^{2\eta})e^{(M-N)\eta}\right]+\cdots, \label{k3Ld-3}\\[4pt]
\Lambda_2(u)|_{u\rightarrow +\infty}&=&-
\frac{1}{4^{2N+3}}e^{4(N+3)u+8\eta}c'\frac{c_1c_2}{c}\left[c\frac{c_1'c_2'}{c'}e^{-2M\eta+\eta}
\right.\no\\[4pt]
&&\left.+(c_1c_2'+c_1'c_2e^{2\eta})e^{(M-N)\eta}\right]+\cdots, \label{k3Ld-4}\\[4pt]
\Lambda_2(u)|_{u\rightarrow -\infty}&=&-
\frac{1}{4^{2N+3}}e^{-4(N+3)u+2\eta}c\frac{c_1'c_2'}{c'}\left[c'\frac{c_1c_2}{c}e^{2M\eta+\eta}\right. \no\\[4pt]
&&\left.+(c_1c_2'+c_1'c_2e^{2\eta})e^{(N-M)\eta}\right]+\cdots.
\label{k3Ld-5}\eea

For the reflection matrix $K^{\pm}$ given by (\ref{K-matrix-2}) and
(\ref{K-matrix-4}), the eigenvalues of the transfer matrices $t(u)$
and $t_2(u)$ are described by the ansatz (\ref{3t1-101}) and
(\ref{3t2-101}) with the decomposition \bea
&&K^{(1)}(u)=e^{2\eta}(e^{-u}\sinh(\zeta+u)-c\sinh 2u) \no\\[4pt]
&&\hspace{2cm}\times (e^{u-\frac {\eta}
2}\sinh(\zeta'-u+\frac{\eta}{2})+c'\sinh 2(u-\frac {\eta} 2)),\\[4pt]
&&K^{(2)}(u)=e^{2\eta}(e^{u+\eta}\sinh(\zeta-u-\eta)+c\sinh
2(u+\eta))\no\\[4pt]
&&\hspace{2cm}\times(e^{-u-\frac {3\eta}
2}\sinh(\zeta'+u+\frac{3\eta}{2})-c'\sinh
2(u+\frac {3\eta} 2)),\\[4pt]
&&K^{(3)}(u)= K^{(2)}(u), \eea and the parameter $h$ takes the
value of
\begin{eqnarray}
&&h=c\frac{c_1'c_2'}{c'}e^{-(M+N)\eta-12\eta}+c'\frac{c_1c_2}{c}e^{(M+N)\eta+16\eta}
-(c_1c_2'+c_1'c_2e^{4\eta}).\label{r3-2}
\end{eqnarray}

For the reflection matrix $K^{\pm}$ given by (\ref{K-matrix-3}) and
(\ref{K-matrix-4}), the eigenvalues of the transfer matrices $t(u)$
and $t_2(u)$ are also characterized by the ansatz (\ref{3t1-101})
and (\ref{3t2-101}) with the parametrization \bea
&&K^{(1)}(u)=e^{2\eta}(e^{u}\sinh(\zeta-u)-ce^{2u}\sinh
2u)\no\\[4pt]
&&\hspace{2cm}\times(e^{-u+\frac {\eta}
2}\sinh(\zeta'+u-\frac{\eta}{2})+c'e^{-2(u-\frac {\eta} 2)}\sinh 2(u-\frac {\eta} 2)), \\[4pt]
&&K^{(2)}(u)=e^{2\eta}(e^{-u-\eta}\sinh(\zeta+u+\eta)+ce^{-2(u+\eta))}\sinh
2(u+\eta))\no\\[4pt]
&&\hspace{2cm}\times(e^{u+\frac {3\eta}
2}\sinh(\zeta'-u-\frac{3\eta}{2})-c'e^{2(u+\frac {3\eta} 2)}\sinh
2(u+\frac {3\eta} 2)),\\[4pt]
&&K^{(3)}(u)= K^{(2)}(u),\eea and the constant $h$ is given by
\begin{eqnarray}
&&h=c\frac{c_1'c_2'}{c'}e^{-(M+N)\eta-11\eta}+c'\frac{c_1c_2}{c}e^{(M+N)\eta+17\eta}
-(c_1c_2'+c_1'c_2e^{2\eta})e^{2\eta}.\label{r3-3}
\end{eqnarray}



\begin{thebibliography}{99}

\bibitem{Bax82} R.J. Baxter, Exactly Solved Models in Statistical Mechanics,
Academic Press, 1982.
\bibitem{Kor93} V.E. Korepin, N. M. Bogoliubov and A. G. Izergin,
Quantum Inverse Scattering Method and Correlation Function,
Cambridge University Press, 1993.

\bibitem{bax1} R.J. Baxter, Phys. Rev. Lett. 26 (1971) 832;\\
R.J. Baxter, Phys. Rev. Lett. 26 (1971) 834;\\
R.J. Baxter, Ann. Phys. 70 (1967) 323.
\bibitem{Tak79} L.A. Takhtadzhan and L. D. Faddeev, Rush. Math.
Surveys 34 (1979) 11.
\bibitem{Skl78} E.K. Sklyanin and L.D. Faddeev, Sov. Phys. Dokl. 23 (1978) 902.
\bibitem{Alc87} F.C. Alcaraz, M.N. Barber, M.T. Batchelor,
R.J. Baxter and G.R.W. Quispel, J. Phys. A 20 (1987) 6397.
\bibitem{Skl88} E.K. Sklyanin, J. Phys. A 21 (1988) 2375.

\bibitem{Cao1} J. Cao, W.-L. Yang, K. Shi and Y. Wang, Phys. Rev. Lett. 111 (2013)
137201.
\bibitem{Cao03} J. Cao, H.-Q. Lin, K.-J. Shi and Y. Wang,
Nucl. Phys. B 663 (2003) 487.

\bibitem{Gie05}J. de Gier and P. Pyatov, J. Stat. Mech.
(2004) P03002; \\
A. Nichols, V. Rittenberg and J. de Gier, J. Stat. Mech. (2005)
P03003; \\
J. de Gier, A. Nichols, P. Pyatov and V. Rittenberg, Nucl. Phys. B
729 (2005) 387.
\bibitem{Gie05-1} J. de Gier and F.H.L. Essler, Phys.
Rev. Lett. 95 (2005) 240601; \\
J. de Gier and F.H.L. Essler, J. Stat. Mech. (2006) P12011.
\bibitem{Baj06} Z. Bajnok, J. Stat. Mech. (2006) P06010.


\bibitem{Bas1} P. Baseilhac, Nucl. Phys. B 754 (2006) 309.
\bibitem{Bas2} P. Baseilhac and K. Koizumi, J. Stat. Mech. (2007)
P09006.
\bibitem{Bas3} P. Baseilhac and S. Belliard, Lett. Math. Phys. 93 (2010)
213; \\
P. Baseilhac and S. Belliard, Nucl. Phys. B 873 (2013) 550.
\bibitem{Bel13} S. Belliard and N. Cramp{\'e}, SIGMA 9 (2013) 072.
\bibitem{Bel15} S. Belliard, Nucl. Phys. B 892 (2015) 1.
\bibitem{Bel15-1} S. Belliard  and R.A. Pimenta, Nucl. Phys. B 894 (2015) 527.
\bibitem{Ava15} J. Avan, S. Belliard, N. Grosjean and R.A. Pimenta, Nucl. Phys. B 899 (2015) 229.

\bibitem{Skl92} E.K. Sklyanin, Lecture Notes in Physics 226 (1985) 196;\\
E.K. Sklyanin, J. Sov. Math. 31 (1985) 3417; \\
E.K. Sklyanin, Prog. Theor. Phys. Suppl. 118 (1995) 35.
\bibitem{Fra08} H. Frahm, A. Seel and T. Wirth, Nucl. Phys. B 802 (2008) 351.
\bibitem{Nic12} G. Niccoli, Nucl. Phys. B 870 (2013) 397; \\
G. Niccoli, J. Phys. A 46 (2013) 075003.
\bibitem{Fad14} S. Faldella, N. Kitanine and G. Niccoli, J. Stat. Mech. (2014) P01011.
\bibitem{Kit14} N. Kitanine, J.-M. Maillet and G. Niccoli, J. Stat. Mech. (2014) P05015.


\bibitem{Nep04} R.I. Nepomechie, J. Phys. A 34
(2001) 9993; \\ R.I. Nepomechie, Nucl. Phys. B 622 (2002) 615; \\
R.I. Nepomechie, J. Stat. Phys. 111 (2003) 1363; \\
R.I. Nepomechie, J. Phys. A 37 (2004) 433.
\bibitem{Yan05} W.-L. Yang and Y.-Z. Zhang, J. High Energy Phys. 12 (2004) 019; \\
W.-L. Yang and Y.-Z. Zhang, J. High Energy Phys. 01 (2005) 021.
\bibitem{Doi06} A. Doikou and P.P. Martins, J. Stat. Mech. (2006)
P06004; \\ A. Doikou, J. Stat. Mech. (2006) P09010.
\bibitem{Yan06} W.-L. Yang, R.I. Nepomechie and Y.-Z. Zhang,
Phys. Lett. B 633 (2006) 664.
\bibitem{Nep13} R.I. Nepomechie, J. Phys. A 46 (2013) 442002.



\bibitem{Cao13NB875}J. Cao, W.-L. Yang, K. Shi and Y. Wang, Nucl. Phys. B 875 (2013) 152.
\bibitem{Cao14NB886}J. Cao, S. Cui, W.-L. Yang, K. Shi and Y. Wang, Nucl. Phys. B 886 (2014) 185.
\bibitem{Cao13NB877}J. Cao, W.-L. Yang, K. Shi and Y. Wang, Nucl. Phys. B 877 (2013) 152.
\bibitem{Cao14JHEP143}J. Cao, W.-L. Yang, K. Shi and Y. Wang, J. High Energy Phys. 04 (2014) 143.
\bibitem{Li14} Y.-Y. Li, J. Cao, W.-L. Yang, K. Shi and Y. Wang, Nucl. Phys. B 884
(2014) 17.
\bibitem{Zha13} X. Zhang, J. Cao, W.-L. Yang, K. Shi and Y. Wang, J. Stat. Mech. (2014) P04031.
\bibitem{Hao14} K. Hao, J. Cao, G.-L. Li, W.-L. Yang, K. Shi and Y. Wang,
J. High Energy Phys. 06 (2014) 128.
\bibitem{Wan15} Y. Wang, W.-L. Yang, J. Cao and K. Shi, Off-Diagonal Bethe Ansatz for Exactly Solvable Models, Springer Press, 2015.

\bibitem{Perk81} J.H.H. Perk and C.L. Schultz, Phys. Lett. A 84 (1981) 407.
\bibitem{Perk83} J.H.H. Perk and C.L. Schultz, Families of commuting transfer matrices in q-state vertex
models, in Non-linear integrable systems - classical theory and
quantum theory, eds. M. Jimbo and T. Miwa, World Scientific, 1983,
pp. 135-152.
\bibitem{Schu83} C.L. Schultz, Physica A 122 (1983) 71.
\bibitem{Perk06} J.H. H. Perk and H. Au-Yang, Yang-Baxter Equation, in Encyclopedia of Mathematical Physics, eds. J.-P. Fran\c{c}oise, G.L. Naber and
T.S. Tsun, Academic Press, 2006.
\bibitem{Baz85} V.V. Bazhanov, Phys. Lett. B 159 (1985) 321.
\bibitem{Jim86} M. Jimbo, Commun. Math. Phys. 102 (1986) 537.

\bibitem{Shik} A. Lima-Santos, Nucl. Phys. B 558 (1999) 637.
\bibitem{Shik2} K. Shi, G. -L. Li, H. Fan and B. Y. Hou, High Energy Phys. and Nucl.
Phys. 24 (2000) 11.
\bibitem{Shik3} R. Malara and A. Lima-Santos, J. Stat. Mech. (2006)
P09013.

\bibitem{Kar79} M. Karowski, Nucl. Phys. B 153 (1979) 244; \\
P.P. Kulish, N.Yu. Reshetikhin and E.K. Sklyanin, Lett. Math. Phys.
5 (1981) 393; \\
P.P. Kulish and E.K. Sklyanin, Lecture Notes in Physics 151 (1982)
61; \\
A.N. Kirillov and N.Yu. Reshetikhin, J. Sov. Math. 35 (1986) 2627;
\\
A.N. Kirillov and N.Yu. Reshetikhin, J. Phys. A 20 (1987) 1565.

\bibitem{Mez92} L. Mezincescu and R.I. Nepomechie, Nucl. Phys. B 372 (1992)
597.
\bibitem{Zho96} Y.-K. Zhou, Nucl. Phys. B 458 (1996) 504.


\bibitem{8Mel054} H.J. de Vega and A. Gonz$\acute{a}$lez-Ruiz, Nucl. Phys. B 417 (1994) 553.

\bibitem{8Mel055} H.J. de Vega and A. Gonz$\acute{a}$lez-Ruiz, Mod. Phys. Lett. A 09 (1994) 2207.

\bibitem{8Mel056} G.-L. Li, R. H. Yue and B. Y. Hou, Nucl. Phys. B 586 (2000)
711.



\end{thebibliography}
\end{document}